# Quantum weak values and logic, an uneasy couple

B. E. Y. Svensson

*Theoretical High Energy Physics,*

*Department of Astronomy and Theoretical Physics,*

*Lund University, Sölvegatan 14, SE-22362 Lund, Sweden*

Abstract

Quantum mechanical weak values of projection operators have been used to answer which-way questions, *e.g.* to trace which arms in a multiple Mach-Zehnder setup a particle may have traversed from a given initial to a prescribed final state. I show that this procedure might lead to logical inconsistencies in the sense that different methods used to answer composite questions, like "Has the particle traversed the way *X* or the way *Y*?" , may result in different answers depending on which methods are used to find the answer. I illustrate the problem by considering some examples: the "quantum pigeonhole" framework of Aharonov *et al*, the three-box problem, and Hardy's paradox. To prepare the ground for my main conclusion on the incompatibility in certain cases of weak values and logic, I study the corresponding situation for strong/projective measurements. In this case, no logical inconsistencies occur provided one is always careful in specifying exactly to which ensemble or sample space one refers.

My results cast doubts on the utility of quantum weak values in treating cases like the examples mentioned.

## 1. Introduction

How do you ascribe a property to a quantum mechanical system? In the framework of standard quantum mechanics (QM) the answer is clear: by establishing (or at least outlining) an experimental procedure to measure it. As is well-known, to elucidate this answer has been a persistent theme in the conceptual analysis of QM from its very early days [1].

New aspects of this problem appear when one explicitly considers quantum systems that are specified not only by the preparation or "preselection", *i.e.* the specification of an initial state $| in >$, but also by the specification of a final state by "postselection" of a state $| f >$. One could then be interested in determining properties of the system as it evolves from the pre- to the postselected state. These would have to be found by suitable intermediate measurements on the system. Although phrased in slightly different terms, this is the view taken in the so called consistent histories approach to QM; see *e.g.* [2, 3] for reviews and further references[1]. It is also the view taken in the approach by Aharonov and his many collaborators, starting with paper [4] (in the sequel referred to as ABL) and continuing in the development of the concepts of weak measurements and weak values [5]; for some reviews see [6 - 12]. In this paper, I shall mostly follow this line of thought of Aharonov and collaborators.

As an illustration, consider a Mach-Zehnder interferometer (MZI) shown schematically in figure 1. Let the preselected state be the incoming state as in the figure. It is split into the two arms of the MZI at the first beam splitter. For the postselected state, choose the state corresponding to the beam particle reaching the detector $D$. In this configuration, the probability for particles to end up in detector $D$ is zero. In turn, this is interpreted as interference involving both arms of the MZI: there is a distinct correlation effect between the two arms when you make no measurement of which arm the particles went through. On the other hand, if you try to make a strong, *i.e.* projective, measurement to find out through which arm the particle went, you will end up with particles hitting the detector $D$: the probability of particles reaching that detector is now non-zero. In sum, the probability for the particles to hit detector $D$ gives information on "which-way" properties of the particles.

The fact that you may get different answers depending on the type of measurement you make on a quantum system is what Feynman [13] – originally for the two-slit setup – referred to when he said that it "has in it the heart of quantum mechanics" and that it "is impossible, absolutely impossible, to explain in any classical way." Griffiths [2, 3] refers to it as the lack of unicity in QM: the properties to be ascribed to a quantum system depend on the way it is measured.

The purpose of this article is to elucidate this type of reasoning from several different points of view. For a pre- and postselected quantum system, I consider what strong as well as weak measurements may reveal of the intermediate state of the system  I shall be specially interested in investigating possible ambiguities when one tries to combine the result of different measurements.

---

[1] One purpose of the consistent histories approach is to get rid of the so-called measurement problem resulting in a slightly different viewpoint compared to conventional presentations of QM, and in the use of a different terminology. However, I hope I do no essential injustice to the consistent histories approach when I stick to the more conventional viewpoint.

I have nothing new to contribute regarding strong measurement, *i.e.* to the original ABL approach [4] or to the consistent histories approach [2, 3]. However, I will discuss these better known cases to provide background and contrast to my main results, which are that logical inconsistencies may arise with regard to weak values as indicating properties of a quantum system.

In this context, several characteristics of a weak value should be noted.

A basic one is that a weak value does indeed involve an incoming, preselected, state $|in>$ as well as a final, postselected state $|f>$ for the system under investigation [5 - 12]. The interest is then focused on what may be said regarding the system at an intermediate time, *i.e.* during the period between the pre- and the postselection. The main new aspect in the weak value approach compared to the original ABL approach [4] (which involves invoking ordinary projective/strong measurements) is that one now invokes weak intermediate measurements. As usually presented [5 - 12], a weak measurement crucially involves a von Neumann type interaction with a measuring device, a meter. It is in the limit of the strength of this measuring interaction tending to zero that a weak value emerges. Among other things, the weakness of the measurement implies that when performing a second weak measurement on the system, one may, to first order in the weak measurement strength, reason as if the first weak measurement had not been performed. Thus, the quantum state of the system will, to that order, effectively remain unchanged under the weak measurement interaction. As will be seen, this will prove pivotal for my arguments in this article.

Still another important feature of a weak value is that it is linearly dependent on the observable under study. This implies an additivity law: the weak value of a sum of operators equals the sum of their weak values. This is true irrespective of whether the operators commute or not. Weak values therefore provide a tool for investigating also non-compatible observables on one and the same system.

An issue of some contention regarding weak values has been its interpretation: what property of the system under investigation does a weak value reveal? I particularly refer to [6 , 7, 10, 14, 15] and references therein for a presentation of this issue and for some arguments *pro* and *con* various opinions.

In the present article, I shall partly sidestep the issue of interpretation by concentrating on weak values of projection operators. Nor shall I be interested in the precise value of their weak values, only whether they vanish or not. The basic assumption – following Vaidman [16] – is that a non-vanishing (vanishing) value of a projector weak value indicates whether the system has intermediately been (respectively not been) in the particular state represented by that projector. A full description of this assumption is given in section 3.4.1 below.

An important element in my argument will be the correspondence rules between the QM formalism and logic (see, *e.g.* [2]). In particular, I will use the rules that the product of two commuting projection operators corresponds to the logical operation *AND* of a conjunction and that the sum of two such operators, the product of which vanishes, corresponds to the logical operation *OR* of a (non-exclusive) disjunction. These rules are well-established for strong/projective measurements. I argue in sections 3.4.3 and 3.4.4 that they apply also to weak values.

Using the aforementioned basic properties of weak values, I will point out that their use might lead to situations with logical inconsistencies: one line of seemingly valid arguments lead to the conclusion that the system has been in a particular set of intermediate states, another line of seemingly equally valid arguments shows the opposite.

In my presentation, I shall use the concept of (coherent) quantum ensembles. These are identically prepared copies of the system under investigation, each copy represented by one and the same state vector in a Hilbert space. A quantum ensemble – in the sequel often simply called an ensemble – is the sample space on which probabilities are defined. Quantum ensembles result from strong/projective measurements.

I use the ensemble notion as a convenient concept to emphasize the trivial but important fact that probabilities for a quantum system, and its statistical properties in general, crucially depend on which ensemble – or sample space – the probabilities refer to. Missing to note which ensemble a certain probability refers to may lead to erroneous conclusions. In fact, this is an important concern of the consistent histories approach [2, 3].

In discussing weak values, and also within the ABL framework, some proponents prefer the so called "two-state-vector formalism" [6]. I will here use a conventional textbook approach, like in [2]. Needless to say, the two different ways of presenting a given physical situation are totally equivalent. There is a unique, one-to-one correspondence: any valid statement in one formalism has a corresponding valid statement in the other.

I will start my presentation in section 2 by discussing a particular example, the three-qubit system introduced by Aharonov *et al.* in what I will call their "pigeonhole paper" [17], see also [18]. I shall not be primarily interested in the problem they set out to tackle – although I will also comment on that – but use their setting with an eight-dimensional Hilbert space as a rich enough arena to illustrate my points.

Next, in section 3, I turn to a setting with a more general quantum system and present my arguments in more detail. I analyze several different cases that might occur when combining two projectors, either by addition – which, as stated, is taken to correspond to the logical operator *OR*– or by multiplication, taken to correspond to the logical operator *AND*. Here, I present my main new results.

 In the following section 4, I illustrate my findings by applying them to some further concrete examples; the so-called three-box problem [19] and Hardy's paradox [20, 21].

In the final section, I summarize my main results and their consequences.

## 2. The pigeonhole framework

### 2.1. The setup

Consider the setup introduced by Aharonov *et al* [17] . They study a three-qubit system with each qubit called a "particle" or, in a more picturesque language, a "pigeon". The two basis states for each qubit are denoted $|L>$ and $|R>$, symbolizing the left and right arm of a Mach-Zehnder interferometer (MZI), or, alternatively, the up and down $\sigma_z$ eigenstates of a spin-½ particle. (In the paper [17], these states are described as each particle/pigeon being in the *L* respectively the *R* "box".) The three-particle system is prepared – preselected – in the direct-product state

$$| in > = | + >_1 \; | + >_2 \; | + >_3 \; , \tag{1}$$

where

$$| + >_1 = ( | L >_1 + | R >_1 ) / \sqrt{2} \; , \tag{2}$$

and similarly for particles *2* and 3; in the usual spin formalism, such a $| + >$- state is the spin up eigenstate of the corresponding $\sigma_x$ .

The authors are ultimately interested in applying an ABL analysis [4], *i.e.* in investigating what can be said about the system at an intermediate time in its evolution from the preselected state to a final, postselected, state $|f>$, assuming there is nothing but free time-evolution between the different preparations/measurements.

For the postselected state the authors choose another direct-product state,

$$|f> = | + i >_1 \; | + i >_2 \; | + i >_3 \; , \tag{3}$$

where

$$| + i >_1 = ( | L >_1 + i | R >_1 ) / \sqrt{2} \; , \tag{4}$$

with a similar notation for particles *2* and *3*; in the usual spin formalism, they are spin up eigenstates of the corresponding $\sigma_y$ .

For the intermediate situation the authors consider different combinations – to be specified shortly – of projection operators

$$\Pi_1^L = | L >_1 \; _1 < L | \; , \tag{5}$$

with a similar definition for the other combinations of "boxes" *L* and *R* and "particles" *1, 2* and *3* ; these are projection operators onto the eigenstates of the corresponding spin operator $\sigma_z$ .

A particular question the authors of [17] analyze is whether, at an intermediate time, any two of the three particles can be in identical MZI arms (or "boxes"). To this end, the authors investigate correlations between different states of the particles in terms of properties of the three-particle system at an intermediate time in its evolution, *i.e.* after the preselection but before the postselection.

Let me however here start by being very elementary and treat a few more pedestrian questions before I comment on the approach of [17].

Also, I shall for the time being restrict myself to considering a two-particle system consisting of the particles *1* and *2* only, not the full three-particle framework.

### 2.2. *No postselection, strong measurement (two-particle states)*

For this two-particle system, the first questions are what may be said of the system without at all making any postselection. Questions to be asked will be related to the (strong) measurement of (combinations of) projection operators like $\Pi_1^L = |L>_{1\ 1}<L|$ of eq (5), in particular to the combinations

$$\Pi_{12}{}^{same} = \Pi_1^L \Pi_2^L + \Pi_1^R \Pi_2^R , \tag{6}$$

and

$$\Pi_{12}{}^{diff} = \Pi_1^L \Pi_2^R + \Pi_1^R \Pi_2^L , \tag{7}$$

which measure two-particle correlations. In other words, one is interested in whether the "pigeons" are in the same "boxes", irrespective of which box, or in different boxes, again irrespective of which one. Note that these two operators form a commuting and complete set of projection operators.

The rules of conventional QM say that strongly measuring these operators on an ensemble of systems in a state $|in> = |+>_1 |+>_2$ (eq. (1) but without the third ket), leads to a split of the preselected quantum ensemble – the one represented by the state $|in>$ – into two new quantum ensembles (let me call them *SAME* and *DIFF*) corresponding to the "collapse" of the $|in>$ -state. In detail, the state $|in>$, now written as

$$|in> = \tfrac{1}{2} ( |L>_1 |L>_2 + |R>_1 |R>_2 + |L>_1 |R>_2 + |R>_1 |L>_2 ) , \tag{8}$$

turns into

$$|same> = 1/\sqrt{2} ( |L>_1 |L>_2 + |R>_1 |R>_2 ) , \tag{9}$$

with probability

$$\text{prob}\ (same \mid in\ ) = <in \mid \Pi_{12}{}^{same} \mid in> =$$

$$= <in \mid \Pi_1{}^L \Pi_2{}^L \mid in> \ + \ <in \mid \Pi_1{}^R \Pi_2{}^R \mid in> =$$

$$= \ \text{prob}\ (L_1\ L_2 \mid in\ ) + \text{prob}\ (R_1\ R_2 \mid in\ ) = \ ½\ , \qquad (10)$$

respectively into

$$\mid diff\ > = 1/\sqrt{2}\ (\ \mid L>_1 \mid R>_2 + \mid R>_1 \mid L>_2)\ , \qquad (11)$$

with probability

$$\text{prob}\ (diff \mid in\ ) = <in \mid \Pi_{12}{}^{diff} \mid in> =$$

$$= <in \mid \Pi_1{}^L \Pi_2{}^R \mid in> \ + \ <in \mid \Pi_1{}^R \Pi_2{}^L \mid in> =$$

$$= \ \text{prob}\ (L_1\ R_2 \mid in\ ) + \text{prob}\ (R_1\ L_2 \mid in\ ) = \ ½. \qquad (12)$$

(Here, prob $(L_1\ L_2 \mid in\ ) = <in \mid \Pi_1{}^L \Pi_2{}^L \mid in>$ is the probability of finding both particles in their *L*-boxes, *etc.* for the similarly denoted probabilities. Also observe that, even if as operators, $\Pi_{12}{}^{same} \neq \mid same><same \mid$ and $\Pi_{12}{}^{diff} \neq \mid diff><diff \mid$ , the matrix elements exhibited here and below are equal. )

Notice that the correlation probabilities prob( *same* $\mid$ *in* ) and prob (*diff* $\mid$ *in* ) are the sums of the two probabilities corresponding to the particles being in definite states: the probability for a correlated state is the sum of the probabilities for the two ways this correlation can manifest itself. This corresponds to the fact that the quantum ensemble *SAME* is decomposed into two non-overlapping sub-ensembles – the corresponding Hilbert subspaces are orthogonal – one characterized by the state $\mid L>_1 \mid L>_2$, the other by the orthogonal state $\mid R>_1 \mid R>_2$. A similar decomposition applies to the quantum ensemble *DIFF*.

Note also my notation for (conditional) probabilities. For example, prob( *same* $\mid$ *in* ) denotes the probability of finding the state $\mid same>$ conditioned on the measurement being made on the incoming state $\mid in>$. The notation thus entails not only references to the observable being measured – in the example, the projector $\Pi_{12}{}^{same}$ – but also reference to the quantum ensemble on which the measurement is performed, in the example the ensemble *IN* represented by the state $\mid in>$.

## 2.3. *Postselection, strong intermediate measurement (two-particle states)*

I am now ready to take up a question considered by Aharonov *et al* [17]: what is the (conditional) probability prob (*same* $\mid f, in$ ), respectively prob (*diff* $\mid f, in$ ), for the intermediate states $\mid same>$, respectively $\mid diff>$, *given* that one also enforces a final state $\mid f>$? The answer is contained in the so called ABL formulae, introduced by Aharonov, Bergmann and Lebowitz [4]. For a given preselected state $\mid in>$, the ABL probability prob ($same \mid f, in$ ) and the probability prob ($f \mid same, in$ ) for the state $\mid same >$ to project onto the postselected state $\mid f >$ are related to each other by Bayes' rule

$$\text{prob}\,(same \mid f, in\,)\, x\, \text{prob}\,(f \mid in) =$$

$$= \text{prob}\,(f \mid same, in\,)\, x\, \text{prob}\,(same \mid in)\,, \qquad (13)$$

with a similar relation between prob ($diff \mid f, in$ ) and prob ($f \mid diff, in$ ). For my treatment, I find it convenient to consider the probabilities prob ($f \mid same, in$ ) and prob ($f \mid diff, in$ ) instead of the original ABL probabilities prob ($same \mid f, in$ ) and prob ($diff \mid f, in$ ). I stress that this is only a matter of convenience and does not in any way involve matters of principle; statements involving prob ($f \mid same, in$ ) and prob ($f \mid diff, in$ ) can unequivocally be translated into the ABL formalism and *vice versa*.

I note that the probabilities prob ($f \mid same, in$ ) and prob ($f \mid diff, in$ ) are the (conditional) probabilities for the state corresponding, respectively, to the quantum ensembles *SAME* and *DIFF* to end up in the state $\mid f >$. The ordinary rules of QM give

$$\text{prob}\,(f \mid same, in\,) = \text{prob}\,(f \mid same\,)\, x\, \text{prob}\,(same \mid in\,) =$$

$$= \mid < f \mid \Pi_{12}{}^{same} \mid in> \mid^2\,, \qquad (14)$$

with a similar expression for the *diff* case.

The postselected state is the expression (3) (but without the third particle) and can be written

$$\mid f > = \tfrac{1}{2}\,\big(\,\mid L >_1 \mid L >_2 \,-\, \mid R >_1 \mid R >_2\,+$$

$$+\;\; i \mid L >_1\, \mid R >_2 + i \mid R >_1\, \mid L >_2\big)\,. \qquad (15)$$

It follows that $< f \mid same > = 0$, so the probability is zero for particles *1* and *2* from the *SAME* ensemble to end up in this postselected state $\mid f >$. In the pigeonized language, the two pigeons *1* and *2* are not found in the same box.

On the other hand, if one asks for the probability

$$\text{prob}\,(f \mid L_1\, L_2, in\,) = \text{prob}\,(f \mid L_1\, L_2\,)\, x\, \text{prob}\,(L_1\, L_2 \mid in\,) =$$

$$= \mid < f \mid \Pi_1{}^{L}\, \Pi_2{}^{L} \mid in> \mid^2 \qquad (16)$$

for the two particles to be together in box *L*, one finds a non-vanishing value. The same is true for the corresponding *R*-case. Thus, there is a non-vanishing probability for the particles to be together in a definite box, either *L* or *R*, despite the fact that there is zero probability for the particles to be together in any of them, as probed by $\Pi_{12}{}^{same}$.

This situation is exactly the same as in the single MZI setup of figure 1 described in the introduction above. In that setup, suppose you postselect on the dark port detector. With no measurement to probe through which arm the particle went – analogous to measuring $\Pi_{12}{}^{same}$ in the pigeonhole setup – you have zero probability for finding particles in that detector. But if you make a which-way (projective) measurement onto one of the arms of the MZI – analogous to measuring the $\Pi_1{}^L \Pi_2{}^L$ or $\Pi_1{}^R \Pi_2{}^R$ separately in the pigeonhole setup – you will find particles in the detector.

Is this perplexing and against common sense? It is up to anyone to decide for themselves! But this is indeed the Feynman "heart-of-quantum-mechanics"-issue [13].

Is this a logical inconsistency? No, since the two situations measure different things – they refer to different quantum ensembles – and (strong) measurements strongly "disturb" the state under investigation. In the pigeonhole case, when you measure the correlation $\Pi_{12}{}^{same}$, you probe the ensemble *SAME* created by the intermediate measurement of $\Pi_{12}{}^{same}$. On the other hand, when you measure $\Pi_1{}^L \Pi_2{}^L$, you probe another ensemble, namely the one created by the intermediate measurement of $\Pi_1{}^L \Pi_2{}^L$. The two cases refer to different measurements which cannot be realized on the same ensemble. In fact, the situation illustrates the typical QM feature that it is not legitimate to assign different properties to a quantum system without being specific on how the properties are obtained, *i.e.* measured. In the consistent histories approach [2,3], this is expressed by the fact that QM does not obey what Griffiths calls the unicity rule, meaning that there is more than one framework in which one may reason in a logically consistent fashion. The situation is sometimes, *e.g.* in [21], referred to as "contrafactual". Anyhow, in the case just treated, there is no question of any logical inconsistency or ambiguity, only a choice of viewpoint or framework corresponding to the choice of ensemble/sample space. As we will see in subsection 2.5 below, the situation is drastically different when one invokes *weak* intermediate measurement.

### 2.4. *Postselection, strong intermediate measurement (three-particle states)*

Before I treat the weak measurement case, however, let me turn to the full three-particle setup and examine another issue that the authors of [17] take up: Assuming the full three-particle preselected and postselected states, (1) respectively (3), can *any* pair of the three pigeons be in the same box at an intermediate time?

The authors of [17] answer this question in the negative. They argue, on symmetry grounds, that if pigeons *1* and *2* are in different boxes, so must pigeons *2* and *3* as well as pigeons *1* and *3*. No two pigeons, the authors claim, could be in the same box, thus violating what the authors call the "pigeonhole principle".

However, this argument and its conclusion are fallacious [22]. To find out, for example, whether pigeons *1* and *2* are together in some box with also pigeons *2* and *3* together in some box, one must interrogate the system whether — with a slight alteration of the earlier notation – the probability prob ($f$ | $same_{12}$ AND $same_{23}$, $in$ ) vanishes or not. The established correspondence rule that the logical AND corresponds to a product of (commuting projection) operators means that one should intermediately measure $\Pi_{12}{}^{same}$ $\Pi_{23}{}^{same}$ . This projector product can be written

$$\Pi_{12}{}^{same} \ \Pi_{23}{}^{same} \ = \ \Pi_1{}^L \ \Pi_2{}^L \ \Pi_3{}^L \ + \ \Pi_1{}^R \ \Pi_2{}^R \ \Pi_3{}^R \ = \ \Pi_{123}{}^{same}, \qquad (17)$$

which also defines the three-particle correlation operator $\Pi_{123}{}^{same}$ . It implies an intermediate ensemble described by the state $1/\sqrt{2} \ ( \ | L >_1 | L >_2 \ | L >_3 +$

$+ \ | R >_1 | R >_2 | R >_3 \ )$. This state does project onto the postselected state $| f >$ of eq (4). So even quantum pigeons may thrive together [22] and there is no violation of any pigeonhole principle as stated in [17].

This situation, that two projection operators by themselves give zero probability (the two non-disjoint ensembles each never ends up in the postselected state ) but that their product has a non-zero probability (the intersection of the two subensembles does end up in the postselected state) is again an example of the perplexing non-classical behavior of QM. But as above the perplexity is in the eye of the beholder. There is no question of a logical inconsistence for the very same reason as previously: it requires different measurements to establish the probability for each of the projection operators as well as for their product – *i.e.*, different ensembles are involved – and such measurements are non-compatible.

## 2.5. Postselection, weak intermediate measurement and weak values[2]

As Aharonov, Albert and Vaidman [5] originally showed – and as has later been exploited by many authors in many different ways [6 – 12] – reducing the strength of the intermediate measurement interaction between the system and the measuring device (here called the "meter') has astonishing and fruitful implications. In particular, in the weak measurement limit, the change of the mean value $< Q >$ of the meter's pointer variable $Q$ and of the mean value $< P >$ of the meter's conjugate momentum are directly proportional to, respectively, the real and the imaginary part of the weak value

$$_f ( A )_{weak} = < f \, | \, A \, | \, in > / < f \, | \, in > \qquad (18)$$

of the studied observable $A$. One should also note – and this will subsequently become important– that, to lowest order in the weak measurement strength, the preselected state $| in >$ is not influenced by the weak measurement interaction: when performing a second weak measurement immediately after a first one, one may, to first order in the weak measurement strength, reason as if the first weak measurement had not been performed.

---

[2] In their paper [17], the authors make no explicit use of weak values, and only refer to weak measurements in the section entitled "Nature of Quantum Interaction: A First Experiment".

In other words, to that order, the initially preselected quantum ensemble remains the same and is not influenced by the weak measurement.

The weak value of any observable is therefore a measureable quantity. However, what meaning it has, in other words how to interpret it – what property of the system under investigation does it reveal? – is a matter of some contention [6, 7, 10, 14, 15]. In this section, I assume that a non-vanishing weak value of a projection operator in the pigeonhole framework means that the pigeon(s) has (have) intermediately been in the corresponding state/box, while a vanishing such value means that the pigeons have not occupied that state. This assumption is discussed in more detail in section 3.4.1 below.

Now to some examples. First, consider the two-particle case with only particles *1* and *2* present, and look at $_f(\Pi_{12}{}^{same})_{weak}$. It vanishes. With the assumption just introduced regarding the meaning of such a weak value, this implies that the pigeons *1* and *2* do not occupy the same boxes, neither *L* nor *R*. This is the same conclusion as was reached for the strong measurement. This is not surprising, since the matrix element in the numerator of the definition of the weak value – eq. (18) with the operator *A* a projection operator – also enters (absolute squared) in the probability (14) of the strong measurement treatment of section 2.3.

A new feature appears when one expresses this weak value in terms of its parts,

$$_f(\Pi_{12}{}^{same})_{weak} =\,_f(\Pi_1{}^L\,\Pi_2{}^L\; +\; \Pi_1{}^R\,\Pi_2{}^R)_{weak} =$$

$$=\; _f(\Pi_1{}^L\,\Pi_2{}^L)_{weak} +\,_f(\Pi_1{}^R\,\Pi_2{}^R)_{weak}\;,\qquad(19)$$

and notices that the two terms individually do not vanish,

$$_f(\Pi_1{}^L\,\Pi_2{}^L)_{weak} = i\,/\,2\; =\; -\,_f(\Pi_1{}^R\,\Pi_2{}^R)_{weak}\;.\qquad(20)$$

At face value, there seems to be a logical puzzle here: there is a signal in the meter for the *1-2* pair to be in the *L-* boxes, as well as a signal for the pair to be in the *R*-boxes, but no signal for the particles to be in the same boxes irrespective of which!

Notice the difference in this case to the situation with a strong intermediate measurement treated in section 2.3. In the *weak* measurements case, one may perform the different (weak) measurements – of $\Pi_{12}{}^{same}$, of $\Pi_1{}^L\,\Pi_2{}^L$ and of $\Pi_1{}^R\,\Pi_2{}^R$ – in sequence one after the other, and on different meters, *without disturbing* the system (to lowest order in the weak measurement interaction). (In fact, the weak measurements need not even be performed after each other on the same exemplar of the system; the fact that the initial ensemble remains intact under the weak measurement implies that one may use a new exemplar of the system for each measurement.) In other words, the three measurements can be performed on one and the same quantum ensemble. Since these three measurements lead to different conclusions, there is certainly a logical conundrum: which property of the two-particle system does the weak values really represent, the total absence of correlations or the presence of separate $L_1L_2$ and $R_1R_2$ relations? I will present a more thorough analysis of this conundrum in section 3.4.3 below.

Finally for this section, let me, for the three-particle case, check the joint pair projection operator $\Pi_{12}{}^{same}\,\Pi_{23}{}^{same}$. One easily finds that its weak value is non-vanishing. Again, this is no surprise since the matrix element that enters here is the same as in the strong measurement approach. So also on the weak measurement procedure, the pigeonhole principle is upheld.

## 3. A more general treatment

### 3.1. The framework

Consider now a more general quantum system $S$ represented in a Hilbert space $\mathcal{H}_S$ in which the (normalized) states $|\,a>$, $|\,b>$, $|\,c>$, ….. represent different channels available to the system. The system could be a composite system, like the multiparticle system used in the pigeonhole framework of section 2. By a "channel" I mean a configuration of the system which allows measurement of the projection operator onto the state representing that configuration, *i.e.* $\Pi_a = |\,a><a\,|$ for the channel $a$ represented by the state $|\,a>$, $\Pi_b = |\,b><b\,|$ for the channel $b$ represented by the state $|\,b>$, *etc.* Examples of channels in this sense are an arm of an MZI, a combination of arms for a system of several MZIs, a particular three-particle state of the pigeonhole framework of section 2, or any of the three boxes in the so called three-box problem [19] to be considered in section 4.1.

The different cases I treat will require different assumptions regarding the orthogonality of the projection operators $\Pi_a$ , $\Pi_b$ , *etc.*, that is whether their product $\Pi_a\,\Pi_b$ vanishes or not. I will also explicitly state whenever I assume that they form a complete set, *i.e.*, that their sum $\Pi_a + \Pi_b + \Pi_c + \ldots$ equals the unit operator in $\mathcal{H}_S$ .

I now go through the different cases equivalent to those treated in the pigeonhole framework of section 2.

### 3.2. No postselection, strong measurement

Let me first assume that all the projection operators are orthogonal and form a complete set; in particular, I assume non-degeneracy among the states. This means that a single strong measurement of a projection operator will result in projection of the preselected state $|\,in>$ onto the subspace of $\mathcal{H}_S$ spanned by one of the states $|\,a>$, $|\,b>$, $|\,c>$, ….. . For a large number of identical measurements on the same preselected state $|\,in>$ this corresponds to a subdivision of the initial quantum ensemble *IN* (the one represented by the preselected state $|\,in>$) into disjoint subensembles like $ENS_a$ for $|\,a>$, $ENS_b$ for $|\,b>$, *etc.*; they corresponds to orthogonal, one-dimensional subspaces of the total

Hilbert space $\mathcal{H}_S$. The probability prob ( $a$ / $in$ ) for the system to end up in the subensemble $ENS_a$ is given by the usual expression

$$\text{prob} ( a \text{ / } in ) = <in \mid \Pi_a \mid in>. \tag{21}$$

In this case, there are additivity laws of the type

$$\text{prob} ( a \text{ } OR \text{ } b \text{/ } in ) = <in \mid \Pi_a + \Pi_b \mid in> =$$

$$= <in \mid \Pi_a \mid in> + <in \mid \Pi_b \mid in> =$$

$$= \text{ prob} ( a \text{ / } in ) + \text{prob} ( b \text{/ } in ). \tag{22}$$

In a similar fashion, one may discuss the product of two projectors. Since this requires some further specifications (and since nothing particularly interesting appears under the assumptions made here), I defer the product case to subsection 3.3.2 .

### 3.3. Postselection, strong intermediate measurement

This is the ABL framework [4], the principles of which were presented in some detail in section 2.3.

### 3.3.1 Measuring the sum of two projectors

To start with, let me as in section 3.2 assume all the projection operators to be orthogonal and to form a complete set.

Then, one could, for example, ask for the probability prob ( $f$ / $a$, $in$ ) for states from the subensemble $ENS_a$ to end up in a postselected state $|f>$. The usual rules give

$$\text{prob} ( f \text{ / } a, in ) = \text{prob} ( f \text{ / } a ) \text{ } x \text{ prob} ( a \text{ / } in ) =$$

$$= | <f \mid \Pi_a \mid in> |^2 . \tag{23}$$

There is no simple additivity rule for the union of subensembles, like $ENS_a$ and $ENS_b$ . This follows from the fact that

$$\text{prob} ( f \text{ / } a \text{ } OR \text{ } b, in ) = | <f \mid \Pi_a + \Pi_b \mid in> |^2 \tag{24}$$

is not simply related to the two subensemble probabilities prob ( $f$ / $a$, $in$ ) and prob ( $f$ / $b$, $in$ ). In particular, for $<f \mid \Pi_a \mid in>$ and $<f \mid \Pi_b \mid in>$ non-vanishing but $<f \mid \Pi_a + \Pi_b \mid in>$ equal to zero – as for the pigeonhole case in section 2.3– this is the Feynman "heart-of-QM"–phenomena mentioned above. Again, it is up to anyone to decide whether this is perplexing or not. But it is no logical inconsistency, since it requires three different strong – and therefore projecting – measurements to arrive at the three probabilities under study. To phrase it differently: in QM the two separately defined subensembles $ENS_a$ and $ENS_b$ may behave differently upon postselection compared to the coherently joined subensemble $ENS_{(a \text{ } OR \text{ } b)}$ .

### 3.3.2 Measuring the product of two projectors

Let me next consider the product of two projection operators. To get an interesting case, these two projectors should be commuting (so that their product is a projector, too) and non-orthogonal (if they were orthogonal, their product would trivially vanish). That the projectors commute means that that there is a degeneracy among the channel states in the Hilbert space $\mathcal{H}_S$. Assume for simplicity that this degeneracy is two-fold. The states I shall be interested in can then be characterized by a double label and may be written $| a_1, b_1 >$, $| a_1, b_2 >$, $| a_2, b_1 >$, $| a_2, b_2 >$, *etc.* Let me focus on the projector product $\Pi_{a_1} \Pi_{b_1}$. In the ensemble language, the degeneracy means that there is overlap between the subensembles $ENS_{a_1}$ (lying in the Hilbert subspace spanned by $| a_1, b_1 >$ and $| a_1, b_2 >$) and $ENS_{b_1}$ (lying in the Hilbert subspace spanned by $| a_1, b_1 >$ and $| a_2, b_1 >$) .

One may then ask for the probability prob $( f / a_1 \text{ AND } b_1, in )$ of finding the system in the intersection of $ENS_{a_1}$ with $ENS_{b_1}$ when it its postselected in the state $| f >$. By the correspondence rule – that a logical *AND* conjunction corresponds to a product of the operators involve – one finds

$$\text{prob} \, ( f / a_1 \text{ AND } b_1, in ) = | < f | \, \Pi_{a_1} \Pi_{b_1} | in > |^2 \, . \tag{25}$$

As above, this has no simple relation to the subensemble probabilities prob $( f / a_1, in )$ and prob $( f / b_1, in )$, and might lead to some bafflement. It could, for instance, happen that both prob $( f / a_1, in )$ and prob $( f / b_1, in )$ are zero but that prob $( f / a_1 \text{ AND } b_1, in )$ is non-zero; the pigeonhole setup with $\Pi_{a_1} = \Pi_{12}{}^{same}$ and $\Pi_{b_1} = \Pi_{23}{}^{same}$ furnishes an example. It could also happen that both prob $( f / a_1, in )$ and prob $( f / b_1, in )$ are non-zero but that prob $( f / a_1 \text{ AND } b_1, in )$ equals zero, a concrete example of which will be given for Hardy's paradox setup [20, 21] in section 4.2 below. These are more examples of typical quantum behavior for which no classical explanation is available. As above, it might be perplexing and against common sense, but it constitutes no logical inconsistency. This is guaranteed by the incompatibility of the required strong measurements, *i.e.* by the intermediate measurements being performed on different ensembles.

### 3.4. Postselection, weak intermediate measurement and weak values

### 3.4.1. The Vaidman "past-of-a-quantum-particle" criterion

The first question to face is what kind of property a weak value represents.

For strong, projective measurements, one may base one's reasoning on established, conventional QM rules, expressed in terms of probabilities, for the connection between the mathematical formalism and experimentally accessible entities. However, for weak values there are no such commonly accepted rules. A weak value, formally being a normalized transition amplitude, is certainly not a probability in any conventional meaning of the probability concept[3]. Neither is it an eigenvalue of any (Hermitian) operator, nor in general a (conventional) expectation value.

A clue to what seems to be an uncontroversial interpretation is based on the observation by Vaidman [16] that a non-vanishing (vanishing) weak value of a projection operator can be interpreted as the presence (respectively non-presence) of the system in the channel corresponding to that projection operator[4]. This is due to the fact that in the definition of a weak value of an operator $A$, one assumes the canonical momentum operator of a meter to be coupled to $A$ in the usual von Neumann way. The triggering of the meter – in the sense that at least one of the mean values $< Q >$ and $< P >$ of the meter pointer variable $Q$ and its canonical momentum $P$ is shifted away from their non-measurement values – will then disclose a non-vanishing weak value in the limit of weak measurement interaction (see section 2.5 above). The argument is now that the measurement of the projection operator $\Pi_a$ can trigger the meter in the sense of having $< Q >$ or $< P >$ deviating from its non-measurement value– *i.e.* showing a nonzero weak value $_f(\Pi_a)_{weak}$ – if and only if the system before postselection has a non-zero probability of having been in the intermediate channel $a$ corresponding to $\Pi_a$, *i.e.* if and only if there is a non-zero overlap of the $a$-channel intermediate state with the postselected one.

A further argument supporting such a point of view is that the nominator of the expression for the weak value of a projection operator, say

$$_f(\Pi_a)_{weak} = <f | \Pi_a | in > / <f | in > , \qquad (26)$$

equals the matrix elements that enters absolute squared in the probability prob ( $f / a$, *in* ) of eq (23) in section 3.3.

These arguments lead to the basic assumption made here regarding the interpretation of a weak value of a channel projection operator: its non-vanishing (vanishing) unambiguously signifies the intermediate presence (respectively absence) of the system in that channel. I will refer to this as a "property" of the system: a non-vanishing weak value $_f(\Pi_a)_{weak}$ is taken to be synonymous to the property that "the system, preselected in the state $| in >$ and postselected in the state $| f >$, has (intermediately) been in channel $a$", sometimes even shortened into "the system is (or was or has been) in channel $a$". Similarly, if $_f(\Pi_a)_{weak} = 0$, the system does not have that property. The triggering (or not) of a channel meter thus reveals whether the system is, intermediately between the pre- and postselection, described by a state which at postselection has (or has not) a non-zero amplitude for that channel state.

---

[3] This is not to say that there have not been attempts to enlarge the concept of a probability to , *e.g.*, complex values and to fit complex weak values into such a scheme [23]. This is however well outside conventional theory, and I do not subscribe to such a view here.

[4] Some further properties of a weak value of a projector need to be fulfilled, *e.g.* it must be a "representative" weak value as treated in [24 - 26]. I assume that this is the case for the weak values of the projection operators considered here.

In sum: a signal (respectively no signal) in a channel weak measurement meter – signifying a non-vanishing (respectively a vanishing) weak value of the channel projector – is operationally interpreted as the presence (respectively absence) of the system in that channel in between the pre- and postselection.

### 3.4.2. Sequence of weak measurements

A further crucial characteristic of weak values is that the weakness of the measurement implies that, when performing a second weak measurement immediately after a first one, one may, to first order in the weak measurement strength, reason as if the first weak measurement had not been performed. In the ensemble language, this means that, to that order, one may reason as if a weak measurement effectively leaves the incoming, preselected ensemble *IN* intact. In other words, a weak intermediate measurement does not cause the *IN* ensemble to be split into subensembles as a strong measurement does. Consequently, one may perform several weak measurements (with different meters), none of them changing the original preselected ensemble. In particular, this applies to successive measurements of channel projection operators. Thus, each of their weak values reveals a property of the system[5].

I have now prepared the ground for discussing whether a set of weak values of channel projection operators gives a consistent picture of the properties – presence or absence in the different channels – of the system under investigation.

### 3.4.3 Weakly measuring the sum of two projectors

Let me first assume that the system's channel states are non-degenerate. Consider two channel projection operators, $\Pi_a$ and $\Pi_b$. They are then orthogonal so that $\Pi_a \, \Pi_b = 0$. Then $\Pi_a + \Pi_b$ is also a projector. It is clear from what is said in section 3.4.1 that this projector tests whether the system has the property of being in the channel *a* OR in the channel *b,* where OR stands for the logical operation of (non-exclusive) disjunction: a meter weakly measuring $\Pi_a + \Pi_b$ tests whether the system has a non-vanishing postselected amplitude for a state in the union of $ENS_a$ and $ENS_b$. This is also in agreement with the conventional correspondence rule between logic and the QM formalism , *viz.*, that sum of operators corresponds to the (non-exclusive) logical operation *OR* (see, *e.g.,* [2]).

Now, think of independently weakly measuring all three projectors $\Pi_a$ , $\Pi_b$ and $\Pi_a + \Pi_b$ on the preselected ensemble *IN* and postselecting on the same final state $|f>$. The following different situations may occur:

I.  All three weak values, $_f(\Pi_a)_{weak}$ , $_f(\Pi_b)_{weak}$ and $_f(\Pi_a + \Pi_b)_{weak}$ , vanish. This is perfectly consistent: the system is neither in the *a*-channel, nor in the *b*-channel.

II. All three weak values are non-zero. This is again consistent with the system now being in both the *a*-channel and in the *b*-channel.

---

[5] If none of them produces a "non-representative" weak value; *c.f.* footnote [4]

III. The two weak values $_f(\Pi_a)_{weak}$ and $_f(\Pi_b)_{weak}$ are non-zero but of opposite signs, so that $_f(\Pi_a + \Pi_b)_{weak} = 0$. The interpretation would be that the system is in both of the channels $a$ and $b$ separately, but cannot be found in either of them, unless, that is, that channel is specified!

Case III here is the real conundrum. It is, of course, due to complete destructive interference between the states in the union of $ENS_a$ and $ENS_b$, states that are independently probed by the separate measurement of $_f(\Pi_a)_{weak}$ and $_f(\Pi_b)_{weak}$; the pigeonhole setup discussed in section 2 above furnishes an example. In contrast to the case of a strong measurement, this is now a real logical puzzle: The whole procedure is formulated in a direct, operational manner using meter readings. The measurements are made without disturbance, on one and the same quantum ensemble. It means that an unequivocal conclusion regarding a property of the system, *i.e.* whether it has been – or has not been – in channel $a$ or in channel $b$, cannot be reached. I see no solution to this dilemma, which is a real logical inconsistency.

### 3.4.4 Weakly measuring the product of two projectors

Let me next consider the same case as in section 3.3.2 with two projectors, $\Pi_{a_1}$ and $\Pi_{b_1}$, being non-orthogonal (their product $\Pi_{a_1}\Pi_{b_1}$ does not vanish) and commuting (to ensure that their product is a projector, too). From the arguments in section 3.4.1 and 3.4.2, it follows that this projector tests whether the postselected system has intermediately been in a state in the intersection of $ENS_a$ and $ENS_b$, *i.e.*, whether the system has been in both the $a_1$-channel *AND* in the $b_1$-channel with AND denoting the usual logical operation of conjunction. This interpretation of a product of projectors as representing the logical operation AND is in accordance with the conventional correspondence rules between logic and QM (see, *e.g.* [2]).

 In this case, it might be interesting to find out what different independent weak measurements of the three projectors $\Pi_{a_1}$, $\Pi_{b_1}$ and $\Pi_{a_1}\Pi_{b_1}$ have to say about the system, again assuming the same postselected state $|f>$ for all three cases. The following situations are those that may occur:

(i) All three weak values, $_f(\Pi_{a_1})_{weak}$, $_f(\Pi_{b_1})_{weak}$ and $_f(\Pi_{a_1}\Pi_{b_1})_{weak}$, vanish. The system is neither in the $a_1$-channel nor in the $b_1$-channel, nor in both the $a_1$-channel *AND* the $b_1$-channel. This is perfectly consistent.

(ii) All three weak values are non-zero. One may unequivocally conclude that the system has been in both the $a_1$-channel *AND* the $b_1$-channel, so again no inconsistency.

(iii) Next, suppose $_f(\Pi_{a_1})_{weak} \neq 0$ and $_f(\Pi_{b_1})_{weak} \neq 0$ while $_f(\Pi_{a_1}\Pi_{b_1})_{weak} = 0$ (remember: I only assume $\Pi_{a_1}\Pi_{b_1} \neq 0$ as an operator). Then, from $_f(\Pi_{a_1}\Pi_{b_1})_{weak} = 0$, one concludes that the system cannot have been both in the $a_1$-channel *AND* in the $b_1$-channel, *i.e.* it would, with ordinary logic, *NOT* be in the $a_1$-channel *OR NOT* be in the $b_1$-channel, in contradiction to the assumption $_f(\Pi_{a_1})_{weak} \neq 0$ and $_f(\Pi_{b_1})_{weak} \neq 0$. This is patently an inconsistency. (An example of this situation is exhibited in Hardy's paradox as treated below in section 4.2.)

(iv)    The next case is $_f(\Pi_{a_1})_{weak} = 0$ and $_f(\Pi_{b_1})_{weak} = 0$ while $_f(\Pi_{a_1}\,\Pi_{b_1})_{weak} \neq 0$. So the system is neither in the channel $a_1$ (since $_f(\Pi_{a_1})_{weak} = 0$ ) nor in the channel $b_1$ (since $_f(\Pi_{b_1})_{weak} = 0$ ) but (since $_f(\Pi_{a_1}\,\Pi_{b_1})_{weak} \neq 0$) in both the $a_1$-channel *AND* the $b_1$-channel. This is also an inconsistency. (In fact, this is the $\Pi_{a_1} = \Pi_{12}{}^{same}$ and $\Pi_{b_1} = \Pi_{23}{}^{same}$ pigeonhole case of section 2.5.)

(v)    Suppose next that $_f(\Pi_{a_1})_{weak} \neq 0$ and $_f(\Pi_{b_1})_{weak} = 0$ while $_f(\Pi_{a_1}\,\Pi_{b_1})_{weak} = 0$. The interpretation is that the system has been in the $a_1$-channel but not in the $b_1$-channel and therefore not jointly in the $a_1$- and $b_1$-channel, consistent with $_f(\Pi_{a_1}\,\Pi_{b_1})_{weak} = 0$, so no inconsistency.

(vi)    A final case is $_f(\Pi_{a_1})_{weak} \neq 0$ and $_f(\Pi_{b_1})_{weak} = 0$ while $_f(\Pi_{a_1}\,\Pi_{b_1})_{weak} \neq 0$. The interpretation is that the system is in the $a_1$-channel but not in the in the $b_1$-channel, but somehow succeeding to be in both the $a_1$-channel *AND* in the $b_1$-channel, which is against normal logic. (A concrete example of this situation is the pigeonhole framework with $\Pi_{a_1} = \Pi_1{}^L\,\Pi_2{}^L$ and $\Pi_{b_1} = \Pi_{23}{}^{same}$ implying $\Pi_{a_1}\,\Pi_{b_1} = \Pi_1{}^L\,\Pi_2{}^L\,\Pi_3{}^L$ .)

In conclusion, there are several situations – the cases (iii), (iv) and (vi) – where there are inconsistencies if one, at an operational level – readings of independent meters – apply ordinary logical rules combined with basic properties of weak values. The implication is that it is unclear which weak value to rely on when ascribing properties of intermediate presence or absence in a certain channel of the system under study, a logically very worrying situation.

## 4. Some further examples

### 4.1. The three-box problem

Consider a one-particle system in which the particle could be in any of three "boxes" *A, B* or *C*, represented by states $|\,A>$, $|\,B>$ and $|\,C>$ in a three-dimensional Hilbert space [6,17, 19]. The boxes correspond to what I call channels. Let the preselected state be

$$|\,in> = (\,|\,A> + |\,B> + |\,C>\,)\,/\sqrt{3} \qquad (27)$$

and the postselected state be

$$|\,f> = (\,|\,A> + |\,B> - |\,C>\,)\,/\sqrt{3}\ . \qquad (28)$$

One is interested in finding which intermediate channel the particle may have occupied, *i.e.* properties of the projectors $\Pi_A = |\,A><A\,|$, $\Pi_B = |\,B><B\,|$ and $\Pi_C = |\,C><C\,|$.

Let me first calculate the relevant strong measurement probabilities prob ($f \mid A$, $in$), *etc.* One finds

$$\text{prob}(f \mid A, in) = \mid <f \mid \ \Pi_A \mid in> \mid^2 = 1/9 = \text{prob}(f \mid B, in) =$$

$$= \text{prob}(f \mid C, in), \tag{29}$$

which is nothing astonishing whatsoever.

[Remark. The situation becomes seemingly more dramatic if one instead considers the ABL probabilities proper. One has

$$\text{prob}(A \mid f, in) =$$

$$= \text{prob}(f \mid A, in) / \{ \text{prob}(f \mid A, in) + \text{prob}(f \mid NOT\,A, in) \} = 1, \tag{30}$$

and similarly

$$\text{prob}(B \mid f, in) = 1, \tag{31}$$

while

$$\text{prob}(C \mid f, in) = 1/5. \tag{32}$$

This seem to imply that the probability of finding the particle in box $A$ $OR$ in box $B$ would equal 2, a blatant inconsistency. However, the logical order is restored when one realizes that different (strong) measurements with different intermediate subensembles, *i.e.*, different sample spaces, are involved.]

Consider next the sum of the projectors. One finds

$$\text{prob}(f \mid A\ OR\ B, in) = \mid <f \mid \Pi_A + \Pi_B \mid in> \mid^2 = 4/9, \tag{33}$$

while

$$\text{prob}(f \mid A\ OR\ C, in) = \mid <f \mid\ \mid \Pi_A + \Pi_C \mid in> \mid^2 = 0. \tag{34}$$

The interpretation of the second equality is that there is no particle in either of the $A$-$OR$ the $C$-boxes despite the fact that both separate probabilities, prob ($f \mid A$, $in$) and prob ($f \mid C$, $in$), are non-zero, implying that the particle could be in any specific one. A perplexing situation, maybe, but nothing but the Feynman "heart-of-QM" phenomena [13] and certainly no logical inconsistency, since different strong measurements with different subensembles/sample spaces are involved.

It is worse for the weak values. Indeed, one easily finds $_f(\Pi_A)_{weak}$ and $_f(\Pi_C)_{weak}$ both to be non-zero, a fact that is interpreted as the possible presence of the particle in one of the boxes $A$ $OR$ $C$. But testing this on $\Pi_A + \Pi_C$ gives a vanishing $_f(\Pi_A + \Pi_C)_{weak}$, so no particle in either of the boxes! Phrased differently, one gets different results depending on the method used to answer the question "Has the particle been in the *A- OR* the C-boxes?" This is a logical inconsistency inherent in the weak value approach.

### 4.2. *Hardy's paradox*

Hardy [20] (see also [21]) considered a two-qubit setup consisting of two Mach-Zehnder interferometers (MZIs), one traversed by an electron, the other by a positron. One arm of the positron's MZI intersects one arm of the electron's MZI so that annihilation occurs in that intersection. The setup is illustrated in figure 2, which also contains the notation I will use. I will be very brief and refer to, *e.g.* [15, 21] for more details.

This setup contains one complication compared to the previous examples I have investigated: there is now a non-trivial evolution of the system due to the effect of the different beamsplitters. This has to be duly taken into account.

As preselected state one chooses the state of the two-particle system just after the annihilation, duly evolved. It can be written in different forms[6]:

$$| in > = \{ |N_p> \ |N_e> + \ i \ |I_p> \ |N_e> + |N_p> \ i \ |I_e> \} \ / \sqrt{3} \ =$$

$$= \{ (- |D_p> + i \ | \ B_p >) \ |D_e> + i \ |D_p> \ |B_e> - 3 \ |B_p> \ |B_e> \} / \sqrt{12} \ =$$

$$= \{ - i \sqrt{2} \ |I_p> \ |D_e> + i \ |D_p> \ |B_e> - 3 \ |B_p> \ |B_e> \} / \sqrt{12} \ =$$

$$= \{ - i \sqrt{2} \ |D_p> |I_e> \ + i \ |B_p> D_e> \ - 3 \ |B_p> \ |B_e> \} / \sqrt{12} \qquad (35)$$

Hardy's paradox originates in the observation that the detector arm state $| D_e >$ solely occurs together with the positron arm state $| I_p >$, indicating that a signal in the detector $D_e$ means that the positron has been in the $I_p$ –arm. It is similar for $I_e$ with respect to $D_p$. From a joint signal in both $D_p$ and $D_e$ one would then expect the particles to have taken the $I_p$ - $I_e$ way, in which case they would have annihilated and would not have been able to reach any detector, creating the paradox.

Let me analyze this situation in the same way as the other examples. Firstly, one specifies the postselected state to be $| f > = | D_p > | D_e >$. Then one calculates the matrix elements that enter into the appropriate probabilities and in the relevant weak values. One finds – now with the simplified notation $\hat{I}_p = | I_p > < I_p |$ *etc.* for the projectors – that $< f \ | \ \hat{I}_p \ | \ in > \neq 0$ and $< f \ | \ \hat{I}_e \ | \ in > \neq 0$ but that[7] $< f \ | \ \hat{I}_p \otimes \hat{I}_e \ | \ in > = 0$. This means that Hardy's paradox is an example of case (iii) of section 3.4.4. As there, the result may be judged perplexing, but there is no logical inconsistency as long as one sticks to strong measurements and their probabilities. But contradictions do occur when one employs weak values, since then only one quantum ensemble is involved.

---

[6] In order to comply with the notation in [15, 20, 21], I have changed arm symbols from *L* and *R* in figure 1 to *I* (for "interacting") and *N* (for "non-interacting") in figure 2 and in the text.

[7] For clarity, I insert a direct product sign, $\otimes$, between any two projection operators here and in the remaining formulae of the present section. Also, a one-particle projection operator should read as a two-particle operator with an identity operator for the other particle implied.

In [21], the authors analyzed the situation by considering weak values for all pairs of projection operators. They found

$$_f(\hat{N}_p \otimes \hat{I}_e)_{weak} = 1 = {}_f(\hat{N}_e \otimes \hat{I}_p)_{weak} \tag{36}$$

and

$$_f(\hat{N}_p \otimes \hat{N}_e)_{weak} = -1, \tag{37}$$

implying

$$_f(\hat{N}_p \otimes \hat{I}_e + \hat{N}_p \otimes \hat{N}_e)_{weak} =$$
$$= {}_f(\hat{N}_p \otimes \{\hat{I}_e + \hat{N}_e\})_{weak} = {}_f(\hat{N}_p)_{weak} = 0. \tag{38}$$

Again, this entails a logical inconsistency: there is a signal for particle pairs in the $N_p$ - $I_e$ arms as well as in the $N_p$ - $N_e$ arms but no signal for particles in the $N_p$ - $I_e$ OR the $N_p$ - $N_e$ arms. The answer to the question "Which way did the particles take?" thus depends on the scheme employed to answer it. Neither scheme has preference over the other since they all refer to one and the same ensemble/sample space.

The argument may also be twisted to show the original paradox. To this end, note that, analogously to eq (38), one has

$$0 = {}_f(\hat{N}_e \otimes \hat{I}_p + \hat{N}_e \otimes \hat{N}_p)_{weak} =$$
$$= {}_f(\hat{N}_e \otimes \{\hat{I}_p + \hat{N}_p\})_{weak} = {}_f(\hat{N}_e)_{weak}. \tag{38}$$

Thus, there is no signal for particles in the $N_e$ - or in the $N_p$ -channel, only in the $I_e$ - or in the $I_p$ -channel, where the particles are supposed to have annihilated.

## 5. Summary and conclusions

A fundamental issue in quantum mechanics (QM) is to understand what its formalism has to say about natural phenomena that are amenable to experimental inquires. The idea of studying systems that are both pre- and postselected [4] has introduced some novelty into this discourse, in particular when it is coupled to the idea of a weak measurement [5], resulting in the concept of a weak value.

While the conceptual and logical consequences of the conventional QM approach have been thoroughly discussed since QM was developed, not the least by the quantum founding fathers themselves [1], a similar analysis has largely been lacking concerning weak values. This article gives my contribution to such a discussion.

Accepting the rules and postulates of conventional QM, all evidence shows that it does not contain logical inconsistencies. Yes, there are some conceived paradoxes, like in the double-slit setup, or in Hardy's intriguing nested Mach-Zehnder arrangement with electrons and positrons. But these are no logical inconsistencies. They "only" constitute deviations from what would be expected from a notion of common sense, essentially based on ideas grounded in classical mechanics.

For weak values of projection operators, this "peaceful coexistence" between the QM formalism and logic seems to be broken. As I have shown in this article, there are cases for which different, seemingly logically inviolable lines of arguments lead to contradictory conclusions. For example, there are quantum systems which, from one line of arguments, seem to have occupied either of two states but which, from another line of arguments, have occupied none of them. I invoked nothing else but the appropriate, well-established QM rules. The culprit, therefore, sits in the weak value.

I based my argument on several important features of the weak value of a projection operator. One feature is that this weak value, following Vaidman [16], via the signal it gives in the meter measuring it, may be used as an indicator of how the system has evolved from the pre- to the postselected state: the vanishing or not a projector weak value has a direct operational meaning in terms of meter readings. Another, for my argument equally important feature is that – contrary to what is the case for a usual strong/projective measurement – different weak measurements can be performed one after the other on the same preselected system state without, to lowest order in the weak measurement strength, "collapsing" the state of the system. This means that you may pose a series of different which-way questions to the system without disturbing it. For example, one may ask whether the system has separately taken either this or that way as well as ask if it has taken either of these ways. As I have shown, there are situations of this kind in which one gets logically contradictory answers.

I have no solution to this conundrum. Certainly, weak values have a role to play as an experimental tool; see [27] for a recent example. But their use in investigating what is perceived as paradoxical situations in QM – the three box problem, Hardy's paradox, *etc.* – must be strongly questioned. Indeed, one would tread a logical quagmire if one draws conclusions regarding the properties of such quantum system from weak values.

**Acknowledgement**

My sincere thanks go to Eliahu Cohen who does not agree with me but who, in a long ongoing and challenging e-mail discussion, has greatly helped me to shape my ideas. I am also grateful to Johan Bijnens for carefully reading and commenting on the manuscript.

**References**

[1]    Wheeler, J.A. and Zurek, W.H. (editors): *Quantum theory and measurement.* Princeton University Press (1983)

[2]    Griffiths, R. B.: *Consistent Quantum Theory*, Cambridge University Press (2002)

[3]    Griffiths, R. B.: A consistent quantum ontology, Studies in History and Philosophy of Modern Physics **44**, 93 (2013)

[4]    Aharonov, Y., Bergmann, P.G. and Lebowitz, J.L.: Time symmetry in the quantum process of measurement, Phys.Rev. **134,** B1410 (1964)

[5]    Aharonov, Y., Albert, D.Z. and Vaidman, L.: How the result of a Measurement of a Component of the Spin of a Spin-½ Particle can Turn Out to be 100, Phys Rev Lett **60,** 1351 (1988)

[6]    Aharonov, Y. and Vaidman, L.: The Two-State Vector Formalism: An Updated Review, Lecture Notes in Physics Vol **734** (Springer, Berlin, 2007,), arXiv:quant-ph 0105101v2

[7]    Aharonov,Y., Popescu, S. and Tollaksen, J.: A time-symmetric formulation of quantum mechanics, Phys Today **63**, Nov 2010, p 27-32 and the ensuing correspondence in Phys Today **64**, May 2011, p 8-9, 62-63 and October 2011, p 8-10

[8]    Kofman, A.G., Ashhab, S. and Nori, F.: Nonperturbative theory of weak pre- and post-selected measurement, Phys. Rep. **520**, 43 (2013)

[9]    Tamir, B. and Cohen, E.: Introduction to weak measurements and weak values, Quanta **2**(1), 7 (http://quanta.ws) (2013)

[10]    Svensson, B.E.Y.: Pedagogical review of quantum measurement theory with an emphasis on weak measurements, Quanta **2**(1), 18 (http://quanta.ws) (2013)

[11]    Dressel, J., Malik, M., Miatto, F. M., Jordan, A.N. and Boyd, R.W.: Understanding quantum weak values: Basics and applications, Rev. Mod. Phys. **86,** 307 (2014)

[12]    Aharonov, Y., Cohen, E. and Elitzur, A. C.: Foundations and applications of weak quantum measurements, Phys.Rev. A **89,** 052105 (2014)

[13]     Feynman, R.P., Leighton, R.B., and Sands, M.: *The Feynman Lectures in Physics,* Volume 3, Section 1–1, Addison–Wesley (1965)

[14]     Kastner, R.E.: Weak values and consistent histories in quantum theory, Studies in History and Philosophy of Modern Physics **35,** 57 (2004)

[15]     Svensson, B.E.Y.: What is a quantum-mechanical "weak value" the value of ?, Found. Phys. **43,** 1193 (2013)

[16]     Vaidman, L.: Past of a quantum particle, Phys.Rev. A **87,** 052104 (2013)

[17]     Aharonov,Y., Colombo, F., Popescu, S., Sabadini, I., Struppa, D.C. and Tollaksen, J.: Quantum violation of the pigeonhole principle and the nature of quantum correlations, Proc Natl Acad Sci USA **113**(3), 532-535  (2016)

[18]     Aharonov, Y. and Cohen, E.: Weak Values and Quantum Nonlocality, arXiv:1504.03797 (2015)

[19]     Aharonov, Y. and Vaidman, L.: Complete description of a quantum system at a given time, J.Phys. A **24,** 2315 (1991)

[20]     Hardy, L.: Quantum Mechanics , Local Realistic Theories, and Lorentz-Invariant Realistic Theories, Phys Rev Lett **68**, 298 (1992)

[21]     Aharonov,Y., Botero, A., Popescu, S., Reznik, R. and Tollaksen, J.: Revisiting Hardy's paradox: counterfactual statements, real measurements, entanglement and weak values,  Phys. Lett. A **301**, 130 (2002)

[22]     Svensson, B.E.Y.: Even quantum pigeons may thrive together. Proc Natl Acad Sci USA, **113** (22), E3052 (2016) and the reply by Aharonov,Y., Colombo, F., Popescu, S., Sabadini, I., Struppa, D.C. and Tollaksen, J.: Quantum violation of the pigeonhole principle , Proc Natl Acad Sci USA **113** (22), E3053  (2016)

[23]     Hofmann, H.: What the complex joint probabilities observed in weak measurements can tell us about quantum physics, arXiv:1303.0078 ) (2013) and references therein

[24]     Svensson, B.E.Y.: Non-representative Quantum Mechanical Weak Values, *Found. Phys.* **45,** 1645 (2015)

[25]     Ben-Israel, A. and Vaidman, L.: Comment on 'Non-representative Quantum Mechanical Weak values' (arXiv:1608.07185) (2016)

[26]     Svensson, B.E.Y.: Reply to Ben-Israel and Vaidman, Comment on 'Non-representative Quantum Mechanical Weak values' (arXiv: 1609.08977) (2016)

[27]     Mahler, D.H., Rozema, L., Fisher, K., Vermeyden, L., Resch, K.J., Wiseman, H.M. and Stenberg, A.: Experimental nonlocal and surreal Bohmian trajectories, Sci Adv **2,** e1501466 (2016)

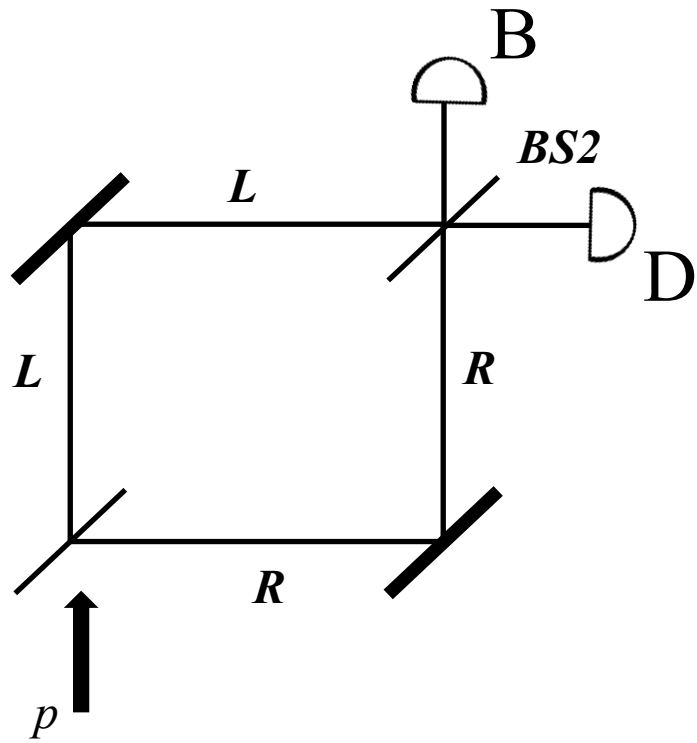

Figure 1. Schematic illustration of a Mach-Zehnder interferometer. The beamsplitters BS1 and BS2 are assumed to be perfect, 50-50 ones. The symbols $B$ (for 'bright') or $D$ (for 'dark') denote detectors.

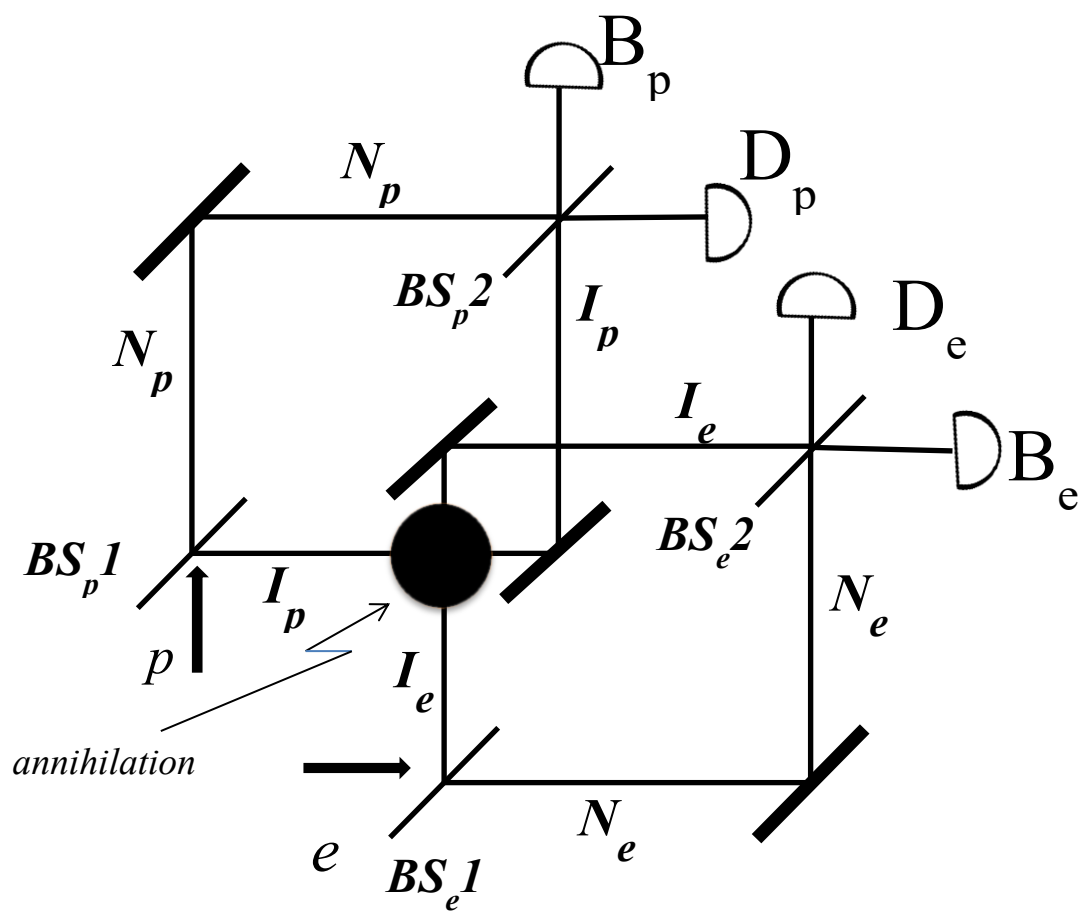

Figure 2. Schematic illustration of the experimental setup for Hardy's paradox [15, 20, 21]. An electron (*e*) and a positron (*p*) each enters its own Mach-Zehnder interferometer with 50-50 beamsplitters (*BS*). The particles are each detected at the respective *B* (for 'bright') or *D* (for 'dark') detectors. They are free to move in the non-interacting arms (*N*) but annihilate as illustrated in the interaction arms (*I*). The paradox is that a pair appears in the *D*-ports, indicating that the particles went through the *I*-arms, even if they should then have annihilated.